\documentclass[twocolumn,prc,nofootinbib,showpacs,epsfig]{revtex4-1}
\usepackage{filecontents}
\usepackage{amsmath}
\usepackage{amssymb}
\usepackage{tabularx}
\usepackage{epsfig}
\usepackage{graphicx}
\usepackage{dcolumn}
\usepackage{array}
\usepackage{bm}
\usepackage{fancyheadings}
\usepackage{longtable}
\usepackage{multirow}
\usepackage{float}
\usepackage{afterpage}
\usepackage{color}

\newcommand{\ra}{\rightarrow }

\newcommand{\be}{\begin{equation}}
\newcommand{\ee}{\end{equation}}
\def\bea{\begin{eqnarray}} 
\def\eea{\end{eqnarray}}

\newcommand{\gess}{\mbox{$^{76}$Ge }}
\newcommand{\asss}{\mbox{$^{76}$As }}
\newcommand{\assf}{\mbox{$^{75}$As }}
\newcommand{\sess}{\mbox{$^{76}$Se }}

\begin{document}
\flushbottom
\title{First observation of the electron capture of \asss}

\author{A. R. Domula}
\author{K. Zuber}
\affiliation{Institut f\"ur Kern- und Teilchenphysik, TU Dresden,  01069 Dresden, Germany}
\today

\begin{abstract}
For a variety of radionuclides which decay via $\beta^{-}$ and a weak $\beta^{+}$-channel the electron-capture (EC) is not observed yet. As the interest of exact decay characteristica  increased again, not least with the need of reliable data for experiments on investigation of the neutrinoless double-beta-decay, an experiment for the investigation of the \asss-EC was performed. The first time observation of the EC of \asss by this experiment resulted in a total branching-ratio for the EC\,/\,$\beta^{+}$-channel of $p_\mathrm{EC}   =   0.0269 \pm \left(0.0080(\mathrm{stat.}) \pm 0.0029(\mathrm{sys.}) \right)$ and $p_\mathrm{EC}   =   0.0263 \pm \left(0.0077(\mathrm{stat.}) \pm 0.0047(\mathrm{sys.}) \right)$ according to two different methods. For the branching of this decay-channel into the first excited- and the ground state of \gess a limit was obtained.
\end{abstract}

\maketitle

\section{Introduction}

The study of weak interaction had important impact in the development of modern particle physics, like the discovery of parity violation and the existence of neutrinos. In order to answer important questions of modern physics, the actual experiments focus on the detection of the neutrinoless double-beta-decay and the search for dark matter. The results of those ever-expanding, technological more and more sophisticated experiments decisively depend on nuclear data and models. One of the few experimental approaches to the creation, verification and improvement of nuclear-models is the precise determination of beta-decay-characteristics. The collection of decay-data is also important for nuclear astrophysics as the detailed path determination of the r- and s-processes bases on half-lives and branching ratios of intermediate nuclei.\\

\begin{figure*}
\centering
\includegraphics[width=0.80\textwidth]{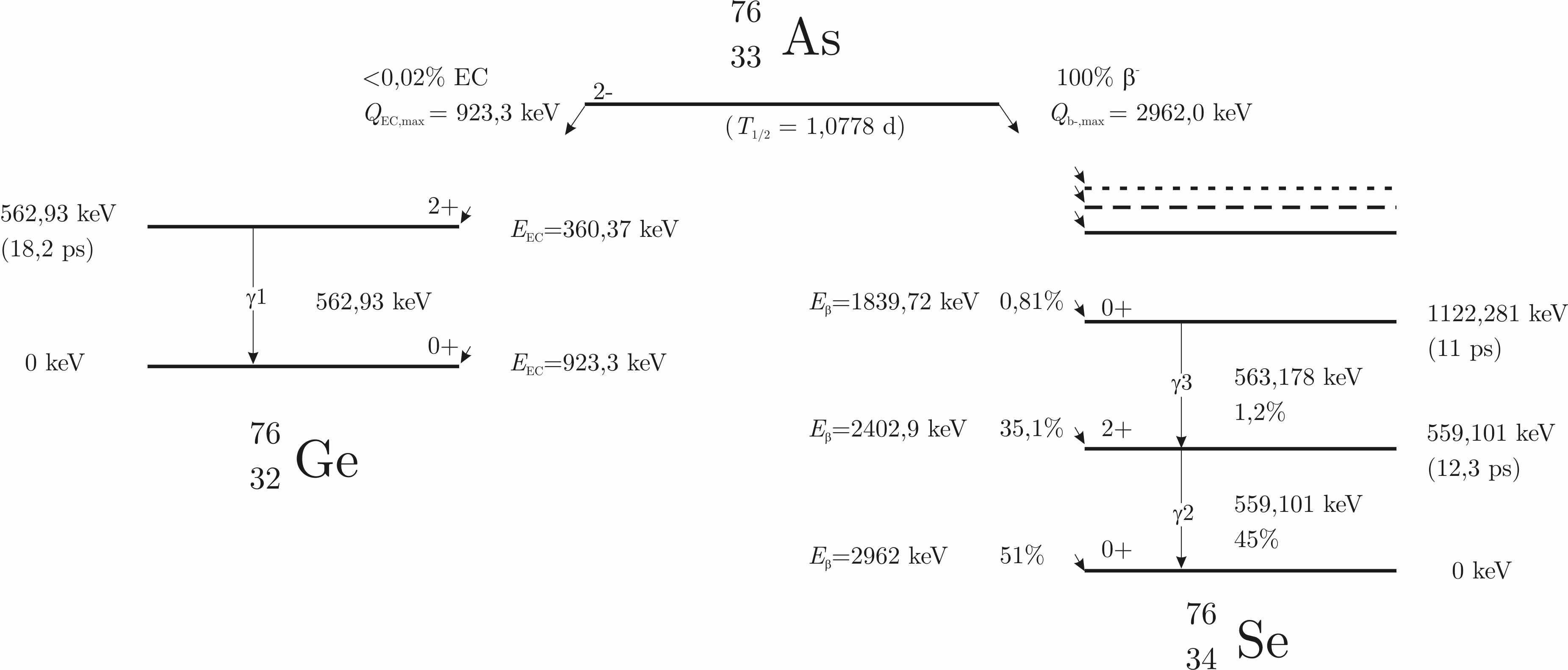}
\caption{ \asss decays with a half life of 1.0778 $\pm$ 0.0020 days via $\beta^-$-decay into \sess with a total branching-ratio of $p_{\beta^{-}}=100\%$. For the weak electron-capture-channel into \gess only a limit is given for its total probability of $p_\mathrm{EC}<0.02\%$. Whereas the $\beta^-$-channel branches with a total probability of 51\,\% into the \sess-ground-state the remaining fraction populates several higher states. Those de-excite via the two lower \sess levels under emission of the $\gamma$-line $\gamma_{2}$ with 559.10\,keV and $\gamma_{3}$ with 563.23\,keV. The first forbidden EC into the 1$^{st}$ excited state of \gess is followed by the emission of a $\gamma_{1}$ with 562.93\,keV, whose energy is nearby $\gamma_{2}$ and $\gamma_{3}$ \cite{singh95}}.
\label{pic:decscheme}
\end{figure*}

In this paper the electron-capture (EC) of \asss towards \gess is studied. Both decay modes, $\beta^{-}$-decay and EC, are energetically possible (Fig. \ref{pic:decscheme}).
The \asss decays into the ground state of both daughters are characterised as
 a $2^- \ra 0^+$ transition with a half-life of 1.0778 $\pm$ 0.0020 days \cite{singh95}. 
The beta decay transition of \asss  $\ra$ \sess with a $Q-$value of 2960.6 $\pm$ 0.9 keV \cite{ame12})  has already been investigated by several experiments ~\cite{mor71,mcm71,singh95,cam98}. A log \textit{ft} = 9.79 has been deduced for the decay into the ground state of \sess \cite{cam98}.
In addition, with a $Q-$value of $Q = 921.5 \pm 0.9$\,keV \cite{ame12} the EC of \asss  is possible into two different levels of the nucleus \gess:
\begin{itemize}
 \item a first forbidden non-unique decay with ($\Delta I ^{\pi_1\pi_2} = 0^-$) into the first excited 2$_1^+$  state at 562.93\,keV 
 \item a first forbidden unique decay with ($\Delta I ^{\pi_1\pi_2} = 2^-$) into the ground state
\end{itemize}
None of these decay-channels have been observed yet. For the branching to the EC-channel various experiments provided upper limits. The oldest experiment, realised by W. Mims and H. Halban \cite{mims51}, was looking for the existence of this channel. An upper limit for the branching ratio 
$r$ of the $\beta^{-}$- with respect to the  EC channel of $r\textless(6.8 \pm 3)\cdot10^{-4}$ has been deduced \cite{mims51}. Using  a proportional-counter J. Scobie \cite{sco57} could not detect any hint for the EC-channel. He obtained an upper limit of 0.02\% for the total branching into the EC-channel, which is still used in todays data evaluation \cite{singh95}. A $\beta$-$\gamma$-coincidence measurement by N.A. Morcos et al. which describes the \asss-decay showed a hint for the first forbidden-EC into the first excited state with a branching-ratio of $p_{1st}^{EC}\,=\,1.3\,\%$ \cite{mor71}. However, on the one hand the value does not match the limit from J. Scobie, on the other hand the signal region which was used by Morcos et al. is enormously dominated with the line at 563.23\,keV which he has dedicated to the $\beta^{-}$-channel in later experiments.

\section{Experimental setup}

Whereas a $\beta^{-}$-/$\beta^{+}$-decay results in an emission of an electron/positron with the corresponding neutrino ($\nu_e/\overline{\nu_e}$), the EC as competing process additionally causes an atomic-shell-vacancy, which is followed by emission of X-rays or Auger-electrons. This fundamental signature is used for the investigation of the \asss-EC. Dominantly a K-shell electron is captured in this process (see Table\,\ref{Table4}) which is followed by Ge-K$_{\alpha}$-X-rays with $E_{x}\,=\,9.88\,$keV ignoring fine structure of capture from other states. Besides the EC to the $\gess$-ground-state the more dominant EC-branch into the first excited 2$^+_1$ state is followed by a $\gamma$-line with $E_{\gamma}\,=\,562.93\,$keV (Figure\,\ref{pic:decscheme}, $\gamma_{1}$).\\

Due to its half life of about one day a sample can easily be measured after an appropriate production process. As arsenic is an anisotopic element the \asss-production via thermal-neutron-capture \assf(n,$\gamma$)\asss with a cross-section of $\sigma_{th}(E_n$\,=\,20\,meV)\,=\,4.6\,barn excludes any other activation products. Pure arsenic is available in a metallic modification, which - by its homogeneity and the loss of parasitic activation of any compound - is preferred as activation
target. Due to the oxidation of arsenic to the toxic As$_2$O$_3$ and its mechanical properties, the handling of a pure arsenic sample in metallic modification for these purposes is quite difficult. Regarding the activation products of any compunds (Table\,\ref{Table1}) the stable oxide of arsenic As$_2$O$_3$ was chosen (99.996\% pure).\\

\begin{figure}[!htb]
\includegraphics[width=0.85\columnwidth]{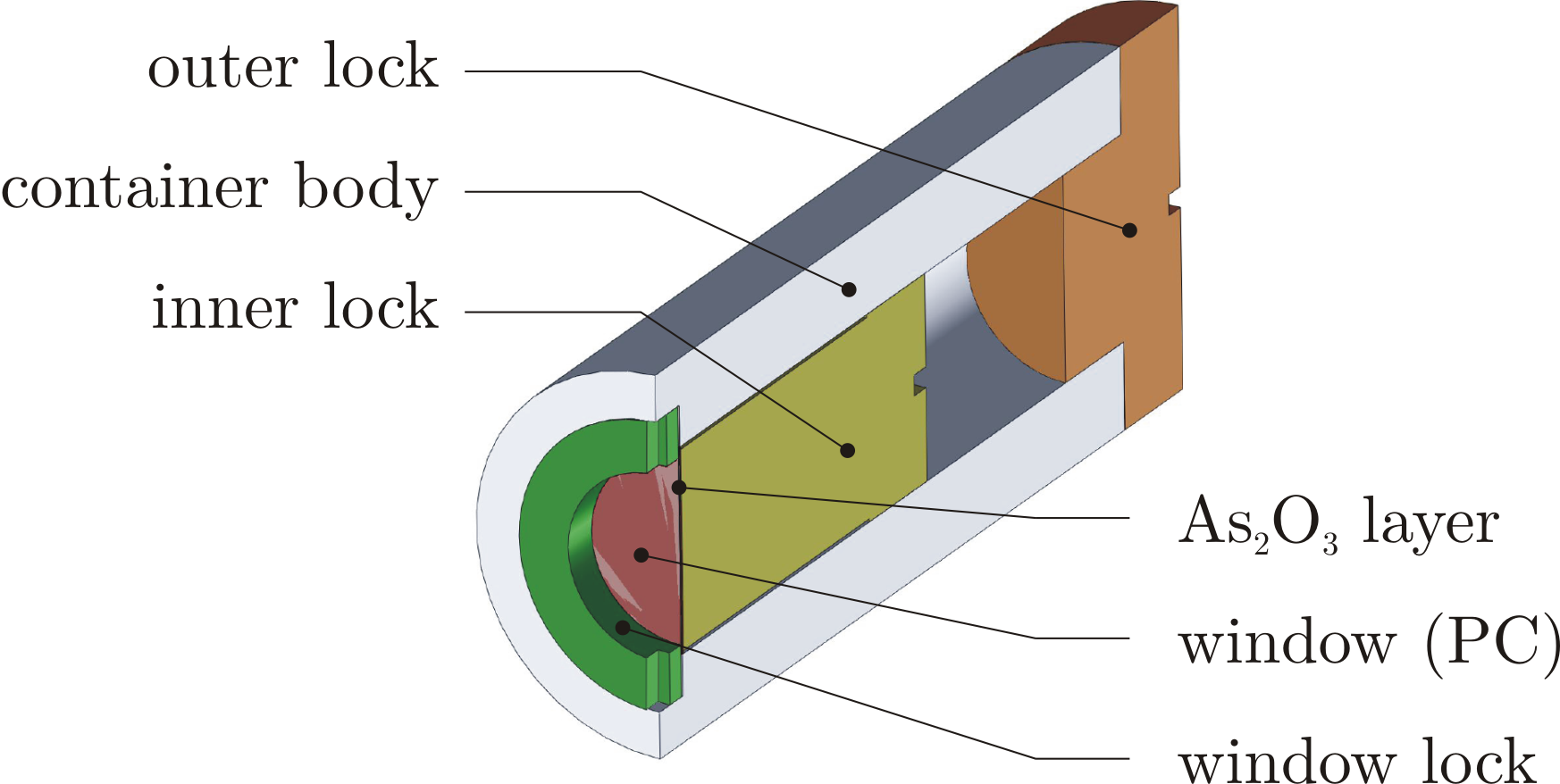}
\caption{The thin As$_{2}$O$_{3}$ layer (red) is fixed on the polycarbonate window-foil with the inner the lock screw (yellow) which allows a well defined, homogeneous shape. All components of the sample container are made of polyethylene.}
\label{fig2}
\end{figure}

For the safe handling of the sample material a container was developed with the special requirements of having
a thin window, with material of a high mechanical strength and a high transmission for the Ge-K$_\alpha$ X-rays so the production of potential parasitic radionuclides is negligible. The powder-sample had to be as thick as necessary for a satisfying counting-statistics and - regarding the self absorption in the sample-material itself - not bigger than reasonable to prevent a large contribution to the background in the X-ray spectra. A container meeting all these requirements was built of polyethylene with a 46\,$\mu$m polycarbonate window (Figure\,\ref{fig2}). A thin layer of 96$\,\mu$m arsenic oxide, which corresponds to the loaded mass of 55\,mg is reasonable small regarding the absorption length for the K$_\alpha$ line of $65.2\,\mu$m ~\cite{xcom12}.\\

The neutron activation was performed at the Department of Energy Technology at TU-Dresden running the 2\,Watt fission reactor AKR-II \cite{akr}. At a thermal-neutron flux of $\varphi_{th}\,=\,4 \times 10^7$\,cm$^{-2}$s$^{-1}$ in the central access pipe of the reactor an activity of \textit{A}($^{76}$As)\,=\,4.4\,kBq was induced during 12 hours irradiation-time.\\

Due to the solely usage of carbon and hydrogen for an arsenic-composite and the construction material almost all of the activation-products are negligible as they require fast neutrons, except $^{19}$O (Table \ref{Table1}). However, due to their short half-life all of possible parasitic activation products decayed during a short sample-cooling phase after irradiation of approx. 2\,h. The amount of the long-living $^{14}$C and $^3$H, which are pure beta emitters with a very low $Q-$value, results in a negligible low activity.

\begin{table}[!htb]
\caption{The potential activation products from neutron reactions with compounds of the sample material and its holder are shown.}
\label{Table1} 
\begin{tabularx}{1.0\linewidth}{p{0.25\linewidth}  p{0.25\linewidth}  p{0.5\linewidth}} \hline \hline
 &   & \\ 
Reaction & $T_{1/2}$  & remarks  \\ 
 &   & \\ 
\hline
 &   & \\ 
$^{16}$O(n,2n)$^{15}$O & 122.24 s & fast neutron reaction \\
$^{16}$O(n,p)$^{16}$N & 7.13 s & fast neutron reaction \\
$^{17}$O(n,p)$^{17}$N & 4.173 s & fast neutron reaction \\
$^{18}$O(n,p)$^{18}$N & 624 ms & fast neutron reaction \\
$^{18}$O(n,$\gamma$)$^{19}$O & 26.91 s & short lived \\

$^{12}$C(n,2n)$^{11}$C & 20.39 ms & fast neutron reaction\\
$^{13}$C(n,$\alpha$)$^{10}$Be & 1.55$\cdot$10$^6$a & fast neutron reaction\\
$^{12}$C(n,p)$^{12}$B & 20.2 ms & fast neutron reaction \\
$^{13}$C(n,p)$^{13}$B & 17.36 ms & fast neutron reaction \\
$^{13}$C(n,$\gamma$)$^{14}$C & 5730 a & long lived, low \textit{E}$_{\beta}$ \\
$^{2}$H(n,$\gamma$)$^{3}$H & 12.33 a & long lived, low \textit{E}$_{\beta}$ \\
 &   & \\ 
\hline
\end{tabularx}
\end{table}

For the sample-measurement a combined X-ray-$\gamma$-detector setup was used: a high resolution Silicon Drift Detector (SDD) for X-rays and an HP-Ge-detector for $\gamma$-counting. The n-type HP-Ge detector with a relative efficiency of 25\%  and a 0.5\,mm Be entrance window (ORTEC GMX20190) was surrounded by a copper cup to suppress Ge-K$_\alpha$-fluorescence produced in the detector itself. For X-ray detection a high resolution 29\,mm$^2$ Silicon Drift Detector (Bruker X-Flash SDD) with an 8\,$\mu$m Be entrance window was used. Both detectors were installed under 90 degrees to each other with a sample distance of 48.8\,mm to the HP-Ge-detector and 5.1\,mm to the SDD. A FAST-Comtec MPA4 ADC-system with a timestamp list-mode was used for data acquisition to establish an X-ray-$\gamma$-coincidence measurement. To control the random coincidences and stability of the peak position of the system, a pulse generator at 25\,Hz has been installed. The pulser peak position was chosen to be in an energy channel equivalent to 3530\,keV, well above the \asss-$Q-$value.\\
Both detectors were characterised for their energy response and full energy efficiency with the pointlike calibration-sources $^{137}$Cs, $^{60}$Co,$^{133}$Ba, $^{152}$Eu and $^{207}$Bi.\\

Once the sample was activated, it cooled down for two hours and was finally counted for a total acquisition time of 329.933\,s, which corresponds to approx. three \asss half-lives. \\

After another cooling of about 2 months were all activation products decayed to a negligible level an XFA (X-ray fluorescence analysis) was performed with the sample container to ensure that the sample material was free of any relevant germanium contamination (Figure\,\ref{pic:X-spec-fl}). The analysis showed no peaks in the Ge-K$_\mathrm{\alpha}$-peak region at $E_\mathrm{x}\,=\,9.88\,$keV and no contaminations in the sample, especially from germanium, which could mimic a comparable signal. From this measurement a model of the fluorescence-background for the later analysis was derived. Moreover an XFA was applied at a pure germanium sample ($99.9999\,\%$ pure) to understand the expected fluorescence-spectra from germanium. Both XFA-measurements were performed in a setup with the same SDD as it was used in the setup described above. For excitation of the sample-material a highly collimated $^{241}$Am source was used. To ensure a very clean photon beam using the   59.6\,keV $\gamma$-line only, an aluminum filter absorbed all lower-energetic components in the photon field provided by the source.\\

\section{Analysis}

The dominant signal for observation of the \asss-EC, the Ge-K$_\alpha$ line at $E_\mathrm{x}\,=\,9.88\,$keV is the spectral main-structure of the analysis. Before the actual sample activation measurement is presented, the analysis of this potential line in the spectra which are retrieved from the activation-sample-measurement the XFA-measurements are discussed shortly.\\

\begin{figure}[!htb]
\includegraphics[width=1.0\columnwidth]{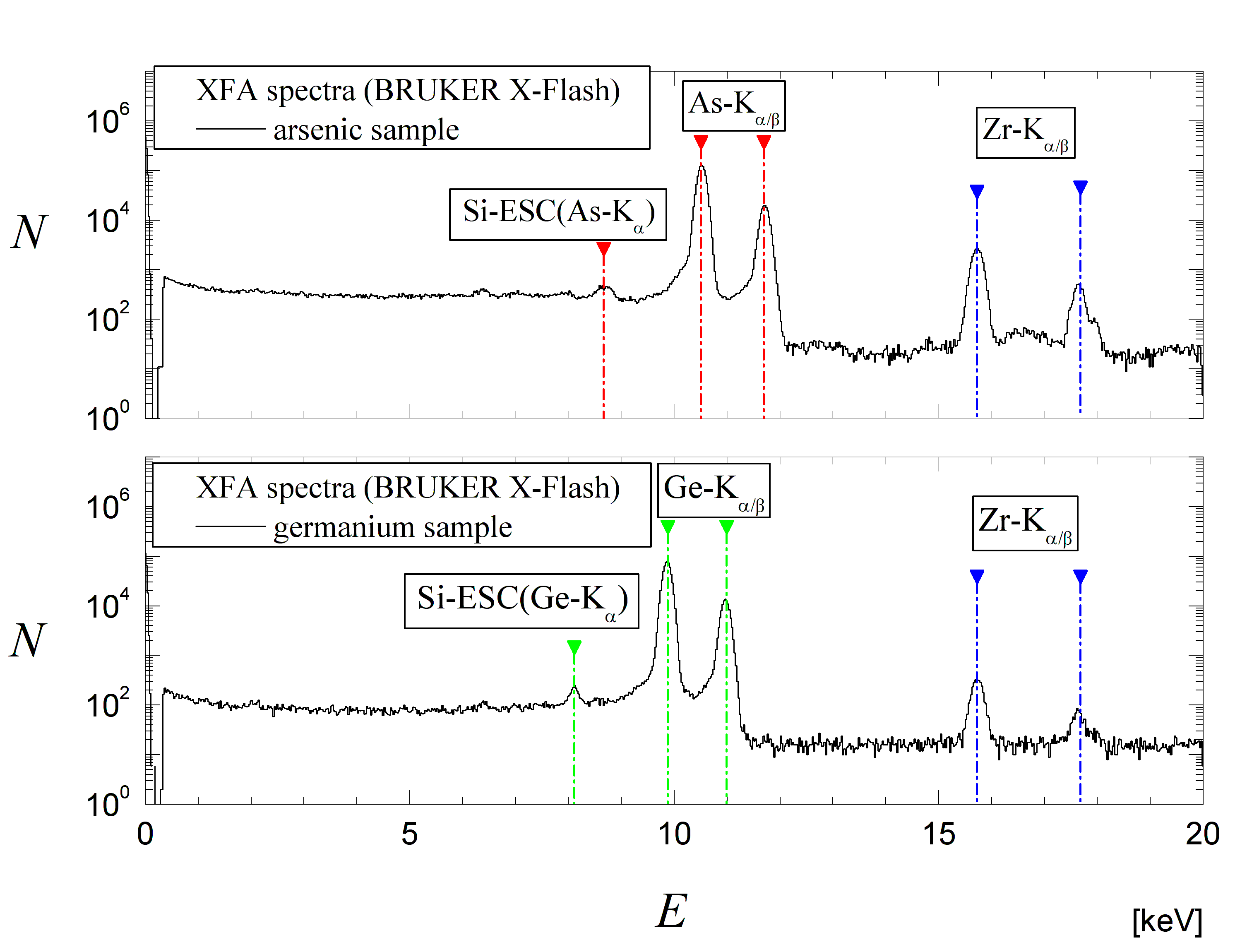}
\caption{\textit{Top:} The XFA-spectra from the arsenic-sample-container is dominated by K$_{\alpha/\beta}$-fluorescence-lines of arsenic (red) from the sample-material itself. The silicon-detector typical Si-escape-peak (blue) at 8.8\,keV does not disturb the region of interest around 9.88\,keV which is not affected by any fluorescence induced background. \textit{Bottom:} The expected signal for the \asss-EC is reflected in the germanium-XFA, which shows the Ge-K$_{\alpha/\beta}$-fluorescence-lines (green) with the corresponding escape-peaks.}
\label{pic:X-spec-fl}
\end{figure}

The XFA of the sample ensured its purity, especially an investigation of impurities which could cause a peak in the region of interest from fluorescence of the sample material itself. The spectrum (Figure\,\ref{pic:X-spec-fl} \textit{top}) shows the strong K$_\mathrm{\alpha/\beta}$-fluorescence-lines of arsenic and zirconium. Arsenic lines have been expected by excitation of the sample-material; the second one occurs from the excitation of the zirconium-collimator of the SDD. Another small peak at $\approx$8.8\,keV was identified as Si-escape-peak (dedicated to As-K$_\alpha/\beta$), which occurs due to a discrete energy excess in the interaction chain of a particle detection: This energy-excess is caused when the incident photon releases a photo-electron in the active-detector-volume, whereas the Si-K$_{\alpha}$-fluorescence-radiation which is followed by the created vacancy escapes and does not contribute to the signal. Mainly the analysis shows that there is no spectral background caused due to fluorescence-excitations in the sample-material itself. Moreover, the fluorescence-spectrum was used to create a background-model-fit for the spectra from the activation sample measurement 
discussed later. In addition, X-ray-peaks show a slight low-energy-tailing \cite{gun77} which exert a small influence at large peaks a gaussian peak-fit-model was chosen, which allows to account for this asymmetry.\\

The XFA-spectrum from the pure germanium sample (Figure\,\ref{pic:X-spec-fl} \textit{bottom}) clearly shows the strong Ge-K$_{\alpha 1/2}$ lines at 9.886 and 9.855\,keV besides the weaker K$_{\beta 1/3}$ lines at 10.982 and 10.975\,keV, which is the expected signal for the \asss-EC.

\begin{figure*}
\centering
\includegraphics[width=0.49\textwidth]{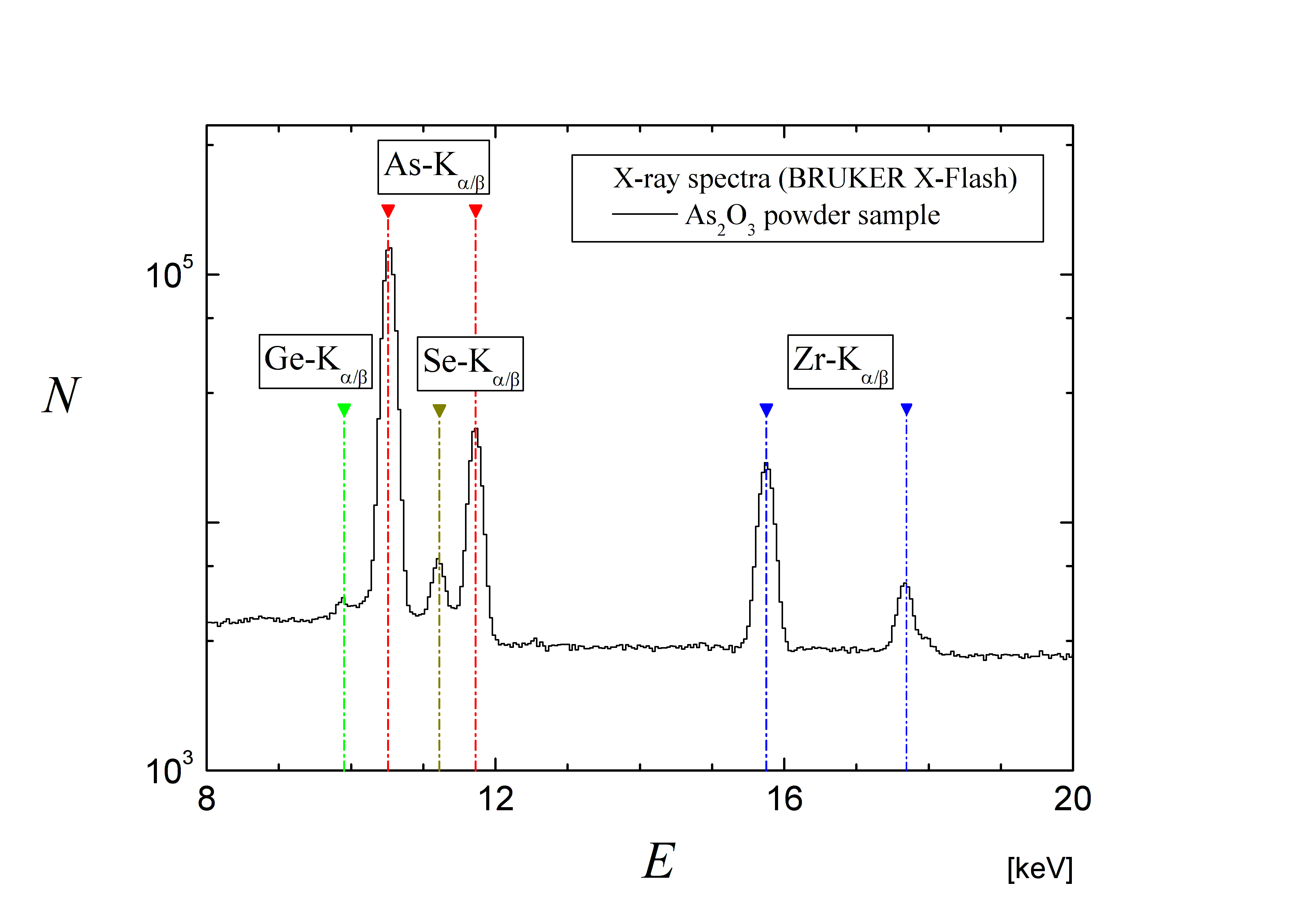}
\includegraphics[width=0.49\textwidth]{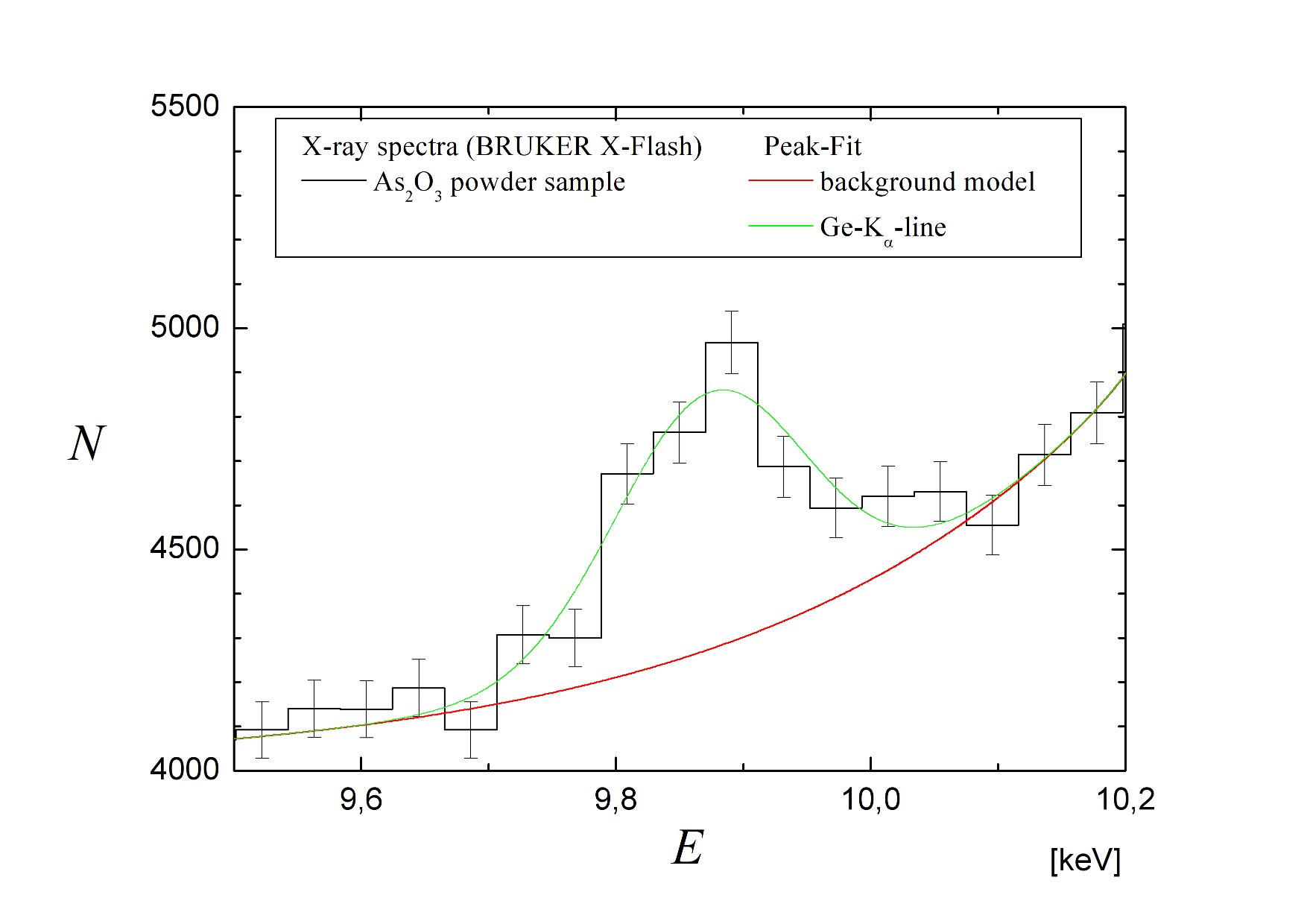}
\caption{\textit{Left:} The X-ray spectra of the activation-sample is dominated by the fluorescence-background from arsenic (red) and zirconium (blue) as expected from the XFA in figure \ref{pic:X-spec-fl}. \textit{Right:} The Ge-K$_{\alpha}$-line at 9.88\,keV as dominant signature of the \asss-EC was analysed with a gaussian fit and the BG-model, which corresponds to the fluorescence-background.}
\label{pic:X-spec}
\end{figure*}

The structures which were described above were finally found in the X-ray spectrum from the activation sample measurement (Figure\,\ref{pic:X-spec} \textit{left}). The measuring background stems almost completely from the dominating $\beta^-$-decay channel of \asss. Corresponding events are classified as either caused by direct interaction of primary ionising particles ($\beta$- and $\gamma$-radiation or X-rays) or by secondary radiation. Latter secondaries like fluorescence radiation, Compton- and photoelectrons or scattered electrons are generated in the interaction of primary particles in the detector- or the sample material itself. The generated Auger-electrons with an energy of $E_e\approx$\,10\,keV have a maximum range of 1.7\,$\mu$m in polycarbonate and won't
 pass the window of the source container.\\

The above discussed XFA-spectra are reflected in the XFA-spectra of the activation-sample (Figure\,\ref{pic:X-spec-fl}): the overlapping Ge-K$_{\alpha 1/2}$-peaks at 9.88\,keV are clearly identified on the lower tail of the As-fluorescence-peaks. With usage of the background parametrisation from the sample container XFA, the Ge- and Se-K$_{\alpha}$-peaks in the X-ray-spectra were fitted using gaussian functions ((Figure\,\ref{pic:X-spec}) \textit{right}, Table~\ref{Table2}). 

\begin{table}[!htb]
\caption{The number of counted peak-events $N$ in the K$_{\alpha1/2}$-lines of germanium and selenium were derived from a gaussian-fit (Figure \ref{pic:X-spec} \textit{right}). For the mean of both overlapping lines $E_\mathrm{X}$ the FEP-efficiencies $\varepsilon_\mathrm{sample}$ is calculated from experimental calibration values with the use of $\kappa1$. The corresponding emission probabilities are taken from \cite{toi98}.}
\label{Table2} 
\begin{tabularx}{1.0\linewidth}{p{0.15\linewidth} p{0.15\linewidth}  p{0.2\linewidth}  p{0.2\linewidth}  p{0.30\linewidth}} \hline \hline
 & &  & & \\ 
$E_\mathrm{X}$ [keV] & Type &  $\nu$ [\%] & $\varepsilon_\mathrm{sample}$ [\%] & \textit{N}  \\ 
 & &  & & \\ 
\hline
 & &  & & \\ 
9.8  & Ge-K$_{\alpha1}$ & $ 31.3\,\textit{11}\,^\dag$    & 1.30 $\textit{21}$ & 2497 $\pm$ 732 \\
     & Ge-K$_{\alpha2}$ & $ 16.1\,\textit{6}\,^\dag$     &  &   \\
 & &  & & \\ 
11.2 & Se-K$_{\alpha1}$ & $0.0316\,\textit{11}\,^\ddag$ & 1.62 $\textit{27}$ & 12977 $\pm$ 664 \\
     & Se-K$_{\alpha2}$ & $0.0163\,\textit{6}\,^\ddag$  & &  \\
 & &  & & \\ 
\hline
 & &  & & \\ 
 \multicolumn{5}{l}{\,\, $\dag$ intensity per 100 K-shell vacancies}\\
 \multicolumn{5}{l}{\,\, $\ddag$ intensity per \asss decay}\\
\end{tabularx}
\end{table}

\begin{table}[!htb]
\caption{Two strong, non-interfering lines from \asss were taken for quantitative analysis. The number of events $N$, the emission-probability $\nu$ from \cite{toi98} the full-energy-efficiency $\varepsilon$ are given for each energy \textit{E}$_\gamma$.}
\label{Table3} 
\begin{tabularx}{1.0\linewidth}{p{0.2\linewidth}  p{0.2\linewidth}  p{0.2\linewidth}  p{0.40\linewidth}} \hline \hline
 &   & & \\ 
$E_\mathrm{\gamma}$ [keV] &  $\nu$ [\%] & $\varepsilon_\gamma$ [\%] & $N$  \\ 
 &   & & \\ 
\hline
 &   & & \\ 
657.041  &  $ 6.2\,\textit{3}$    & 0.569 $\textit{25}$ &619254 $\pm$ 923  \\
 &   & & \\ 
1228.60  &  $ 1.21\,\textit{9}$ & 0.356 $\textit{20}$ & 73212 $\pm$ 396 \\
 &   & & \\ 
\hline
\end{tabularx}
\end{table}

The pulse height-spectrum of the HP-Ge-detector was also analysed to find any hints for impurities. All of the $\gamma$-lines were either dedicated to the \asss-$\beta^-$-decay, to the natural background caused by primordial nuclides like $^{40}$K, $^{235/238}$U and $^{232}$Th decay-chains or to $^{24}$Na. The latter is rarely produced by neutron-capture $^{23}$Na(n,$\gamma$)$^{24}$Na on traces of sodium. Furthermore the $\gamma$-analysis showed no hints for any other impurities.\\

Using the measured data there are two possible ways for the determination of the total EC-$\beta^⁻$ branching-ratio:
\begin{itemize}
 \item based on the ratio of the Se- and Ge-K$_{\alpha}$-lines (discussed in section \ref{ana1})
 \item based on the total activity calculated from the analysis of the gamma spectra (discussed in section \ref{ana2} )
\end{itemize}

\subsection{Monte-Carlo simulations}

The setup was simulated with the Monte-Carlo-Code AMOS~\cite{henn05} to define conversion factors for the determination of the full energy efficiency. These relate the calibration values with point-like geometry to the real sample geometry.

\begin{figure}[!htb]
\includegraphics[width=0.95\columnwidth]{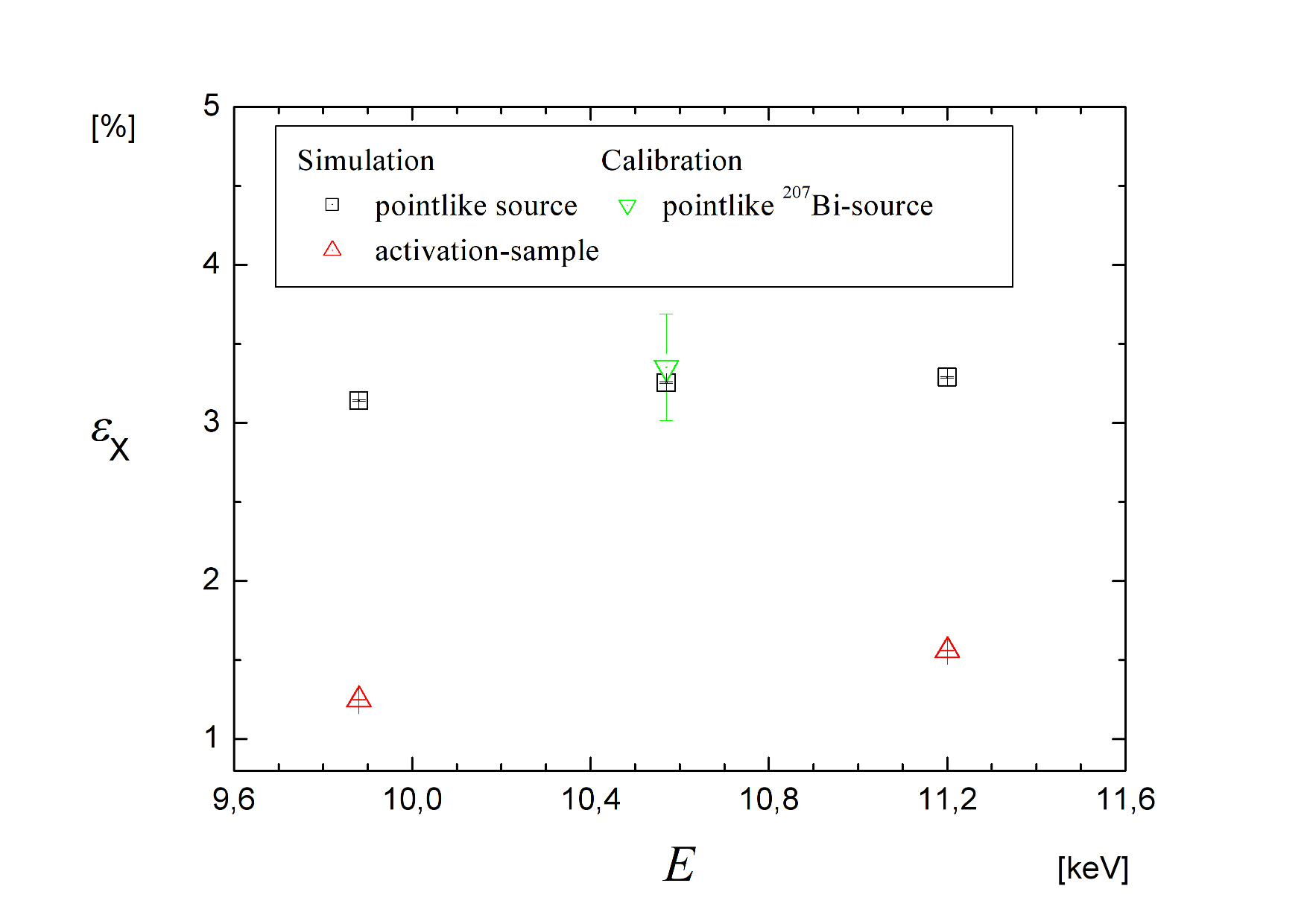}
\caption{The full-energy-efficiency at the Se- and Ge-K$_{\alpha}$-energies were calculated with conversion factors based on Monte-Carlo simulations with the use of the only experimental accessible Bi-L$_{\alpha}$-line from a calibration at 10.57\,keV.} 
\label{pic:MC1}
\end{figure}

The conversion factor

\begin{eqnarray}
\kappa_1 & = &  \frac{\varepsilon_{\mathrm{Sample}}^{\mathrm{MC}}\left(E_{\mathrm{Ge-K_{\alpha}}}\right)}{\varepsilon_{\mathrm{calibration}}^{\mathrm{MC}}\left(E_{\mathrm{Bi-L_{\alpha}}}\right)} =  0.3875\,\,\textit{45}
\label{eq1}
\end{eqnarray}
relates the full energy-efficiencies of the SDD for the energies of the Ge-K$_{\mathrm{\alpha}}$-line at 9.88\,keV and the Bi-L$_{\mathrm{\alpha}}$-line at 10.57\,keV which was accessible from calibration (Figure~\ref{pic:MC1}).\\

A second conversion factor

\begin{eqnarray}
\kappa_2 & = &  \frac{\varepsilon_{\mathrm{Sample}}^{\mathrm{MC}}\left(E_{\mathrm{Ge-K_{\alpha}}}\right)}{\varepsilon_{\mathrm{Sample}}^{\mathrm{MC}}\left(E_{\mathrm{Se-K_{\alpha}}}\right)} = 0.802\,\,\textit{12}
\label{eq2}
\end{eqnarray}

relates the SDD-full energy-efficiencies of the Se- and Ge-K$_{\alpha}$-lines at 11.2\,keV and 9.88\,keV. Both factors are used for further analysis. As the FEP-value from the MC match the calibration value, the MC was validated for the energy range around 10\,keV.

\subsection{Analysis based on the Se-K$_{\alpha}$-line}
\label{ana1}

The event counting in a photo-peak at the energy \textit{E} with an X-ray detector of the full-energy-efficiency $\varepsilon(E)$ and a sample with the specific activity $A_\mathrm{S}$ is given by

\begin{eqnarray}
N & = & A_\mathrm{S}\cdot\varepsilon(E)\cdot\nu(E) \nonumber \\
 &  & \cdot\left( \left(\frac{1}{2}\right)^{\frac{t_\mathrm{D}}{T_{1/2}}} - \left(\frac{1}{2}\right)^{\frac{t_\mathrm{L}+t_\mathrm{D}}{T_{1/2}}}\right)\cdot\frac{T_{1/2}}{ln(2)}
\label{eq3}
\end{eqnarray}

assuming the lifetime $t_\mathrm{L}$, the start of measurement after end of irradation $t_\mathrm{D}$ and the half-life $T_{1/2}$ of the nuclide. 
From equation \ref{eq3} a ratio for the counting in the Ge- and Se-K$_{\alpha}$-line is deduced as

\begin{eqnarray}
\frac{\nu_\mathrm{Ge-K_{\alpha}}\left(^{76}\mathrm{As}\right)}{\nu_\mathrm{Se-K_{\alpha}}\left(^{76}\mathrm{As}\right)\cdot(1-p_\mathrm{EC})} & = & \frac{N_\mathrm{Ge-K_{\alpha}}}{N_\mathrm{Se-K_{\alpha}}}\cdot
\frac{\varepsilon_\mathrm{eff}(11.2\,\mathrm{keV})}{\varepsilon_\mathrm{eff}(9.88\,\mathrm{keV})} \nonumber \\
 & = & \frac{N_\mathrm{Ge-K_{\alpha}}}{N_\mathrm{Se-K_{\alpha}}}\cdot \frac{1}{\kappa_2} 
\label{eq4}
\end{eqnarray}

using $\kappa_2$ from the MC-simulations.\\

The factor (1-\textit{p}$_\mathrm{EC}$) accounts for the different definitions of the X-ray emission-probabilities $\nu_\mathrm{Se-K_\mathrm{\alpha}}$ and $\nu_\mathrm{Ge-K_\mathrm{\alpha}}$: Whereas the emission probability 

\begin{equation}
\nu_\mathrm{Se-K_\mathrm{\alpha}}=\frac{Z_\mathrm{Se-K_{\alpha}}}{Z_\mathrm{\beta^-}}
\label{eq{5}}
\end{equation}
which is deduced from the coefficients for inner conversion and the fluorescence intensities is attached to the number of $\beta^-$-decays $Z_{\beta^-}$ taken from \cite{toi98}, as this is the only evaluated decay channel. The Ge-K$_{\alpha}$-intensity

\begin{equation}
\nu_\mathrm{Ge-K_{\alpha}}=\frac{Z_\mathrm{Ge-K_{\alpha}}}{Z_\mathrm{total}},
\label{eq{6}}
\end{equation}

which is not part of the \asss-decay evaluation yet, is attached to the total number of \asss-decays \textit{Z}$_\mathrm{total}$. The number of Ge- and Se-K$_{\alpha}$-emissions is given by $Z_\mathrm{Ge-K_{\alpha}}$ and $Z_\mathrm{Se-K_{\alpha}}$.

In this notation, the Se-K$_{\alpha}$-emission probability is written as

\begin{eqnarray}
\nu_\mathrm{Se-K_{\alpha}} & = & \frac{Z_\mathrm{Se-K_{\alpha}}}{Z_\mathrm{\beta^-}}   =  \frac{Z_\mathrm{Se-K_{\alpha}}}{Z_\mathrm{total}\cdot p_\mathrm{\beta^-}}  \nonumber\\
 & = & \frac{Z_\mathrm{Se-K_{\alpha}}}{Z_\mathrm{total}\cdot (1-p_\mathrm{EC})}
\label{eq{7}}
\end{eqnarray}

with the total number of decays

\begin{eqnarray}
Z_\mathrm{total} & = & {Z_\mathrm{\beta^-}} + {Z_\mathrm{EC}} \nonumber \\
 & = & Z_\mathrm{total} \cdot({p_\mathrm{\beta^-}} + {p_\mathrm{EC}})
\label{eq{8}}
\end{eqnarray}

The intensity of the Ge-K$_{\alpha}$-line for the \asss-decay

\begin{eqnarray}
\nu_\mathrm{Ge-K_{\alpha}} & = & p_\mathrm{EC}\cdot\omega_k\left(\gess\right)\cdot\left(\nu_\mathrm{\alpha1}+\nu_\mathrm{\alpha2}\right)
\label{eq9}
\end{eqnarray}

is given by the product of the total branching ratio of the EC-channel $p_\mathrm{EC}$ and the fluorescense-intensities $\nu_\mathrm{\alpha1}$ and $\nu_\mathrm{\alpha2}$. The K-capture probability $\omega_\mathrm{K}\left(\gess\right)$ depends on the mass-difference of the initial and final nucleus, the radial wavefunction-component of the captured electron and the energy of the final state. Non-unique forbidden transitions as \asss show an additional dependance on the nuclear-matrix-element but for the K-, L-, M-fractions themselves the same values are assumed as in the allowed case \cite{hanson68}. The determination of the K-fractions for the two \asss-EC-channels with the log-ft code \cite{logft01} showed similar results (Table\,\ref{Table4}).

\begin{table}[!htb]
\caption{The capture-fractions $\omega_\mathrm{i}$ for the \asss-capture at the shell $i$ $(i=\mathrm{K, L, M})$ were determined using \cite{logft01}.}
\label{Table4} 
\begin{tabularx}{1.0\linewidth}{p{0.2\linewidth}  p{0.4\linewidth}  p{0.4\linewidth}} \hline \hline
 &   &  \\ 
 & Non-unique  & 1$^{st}$-unique \\ 
 &   &  \\ 
\hline
 &   &  \\ 
$\omega_\mathrm{K}$ & 0.8773 & 0.865 \\
$\omega_\mathrm{L}$ & 0.10328 $\pm$ 2$\cdot10^{-5}$ & 0.11364 $\pm$ 6$\cdot10^{-5}$ \\
$\omega_\mathrm{M}$ & 0.01947 $\pm$ 3$\cdot10^{-6}$ & 0.0244 $\pm$ 1$\cdot10^{-5}$ \\
 &   &  \\ 
\hline
\end{tabularx}
\end{table}
The branching-ratio of the EC-channel itself is necessary for the determination of the total K-capture-probability from the values in Table\,\ref{Table4}. Currently, these values are unknown, hence an effective $\omega_{\mathrm{K}}^{\mathrm{eff}}$\,=\,(0.8773$\pm$0.0062)\% was determined from the mean values with an error deduced from the deviation of the mean. The total branching ratio 

\begin{eqnarray}
p_\mathrm{EC}  & = & \frac{\chi}{1+\chi}
\label{eq10}
\end{eqnarray}

results from equations \ref{eq4} and \ref{eq9} in

\begin{eqnarray}
\chi  & = & \frac{N_\mathrm{Ge-K_{\alpha}}}{N_\mathrm{Se-K_{\alpha}}}\cdot\frac{1}{\kappa_2}\cdot\frac{\nu_\mathrm{Se-K_{\alpha}}}{\omega_\mathrm{K}\left(\gess\right)\cdot\left(\nu_\mathrm{\alpha1}+\nu_{\alpha2}\right)}
\label{eq11}
\end{eqnarray}

to be

\begin{eqnarray}
\label{totalbr}
p_\mathrm{EC}  & = & 0.0269 \pm \left(0.0080(\mathrm{stat.}) + 0.0029(\mathrm{sys.}) \right).
\label{eq12}
\end{eqnarray}

\subsection{Analysis based on the \asss activity}
\label{ana2}
In analogy to the analysis above the total branching-ratio \textit{p}$_\mathrm{EC}$ can also be determined from the counting in the Ge-K$_{\alpha}$- and an adequate $\gamma$-line from the $\beta^-$-channel (Table \ref{Table3}). In this case equation \ref{eq4} changes to

\begin{eqnarray}
\frac{\nu_{Ge-\mathrm{K_{\alpha}}}\left(^{76}\mathrm{As}\right)}{\nu_{\gamma_i}\left(^{76}\mathrm{As}\right)\cdot(1-p_\mathrm{EC})}  & = & \frac{N_\mathrm{Ge-K_{\alpha}}}{N_{\gamma_i}}\cdot\frac{\varepsilon_\mathrm{eff}(E_{\gamma_i})}{\varepsilon_\mathrm{eff}(E_\mathrm{Ge-K_{\alpha}})}  \nonumber \\
 & & \cdot\nu(\gamma_i)\cdot\left(1-p_\mathrm{EC}\right) \\
 \frac{\nu_{Ge-\mathrm{K_{\alpha}}}\left(^{76}\mathrm{As}\right)}{\nu_{\gamma_i}\left(^{76}\mathrm{As}\right)\cdot(1-p_\mathrm{EC})}  & = & \frac{N_\mathrm{Ge-K_{\alpha}}}{N_{\gamma_i}}\cdot\frac{\varepsilon_\mathrm{eff}(E_{\gamma_i})}{\varepsilon_\mathrm{calibration}(E_\mathrm{Bi-L_{\alpha}})}  \nonumber \\
 & & \cdot\frac{\nu(\gamma_i)\cdot\left(1-p_\mathrm{EC}\right)}{\kappa_1}
 \label{eq20}
\end{eqnarray}

with the use of the conversion factor $\kappa_1$. The total branching ratio (equation \ref{eq10}) results from equations \ref{eq20} and \ref{eq4} in

\begin{eqnarray}
\chi  = \frac{N_{Ge-K_{\alpha}}}{N_{Se-K_{\alpha}}} \cdot \frac{\varepsilon_\mathrm{eff}(E_{\gamma_i})}{\varepsilon_\mathrm{calibration}(E_\mathrm{Bi-L_{\alpha}})\cdot\kappa_1} \nonumber \\
       \cdot \frac{\nu_\mathrm{Se-K_{\alpha}}}{\omega_\mathrm{K}\left(\gess\right)\cdot\left(\nu_\mathrm{\alpha1}+\nu_\mathrm{\alpha2}\right)}
\end{eqnarray}

to be

\begin{equation}
p_\mathrm{EC}   =   0.0263 \pm \left(0.0077(\mathrm{stat.}) + 0.0047(\mathrm{sys.}) \right).
\end{equation}

\subsection{Investigation of the branching of the EC-channel}

As the total $\beta^{+}$/$\beta^{-}$ branching-ratio was determined in sections \ref{ana1} and \ref{ana2} the $\beta^{+}$-channel itself branches with a first forbidden non-unique EC into the first excited state of \gess and with a first forbidden unique channel into the \gess ground-state. Thus the \gess-nucleus de-excites by emission of a 592.93\,keV $\gamma$-ray, the branching of the $\beta^{+}$-channel itself is accessible via X-ray-$\gamma$-coincidence analysis.\\

\begin{figure*}[!htb]
\includegraphics[width=0.95\columnwidth]{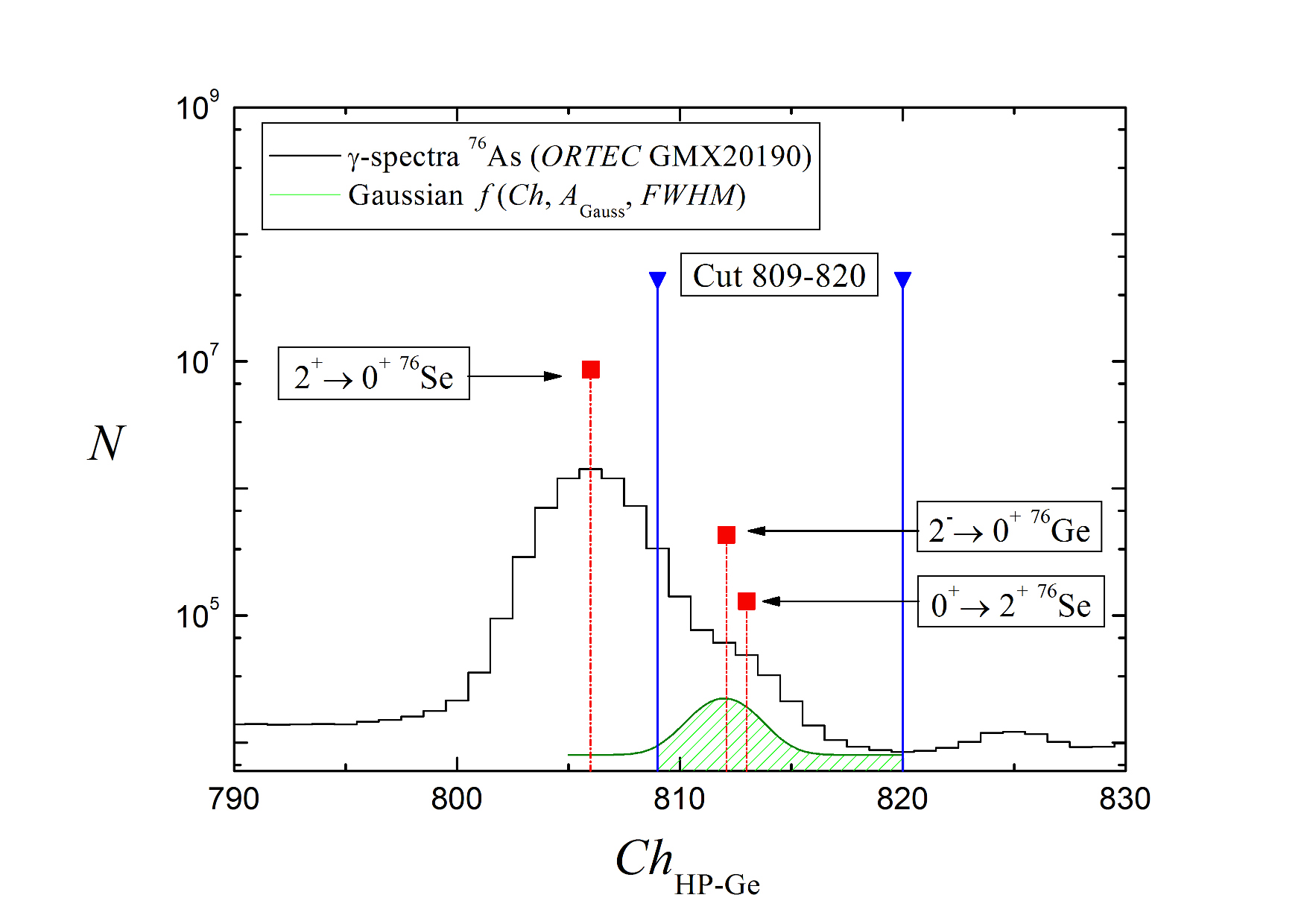}
\includegraphics[width=0.95\columnwidth]{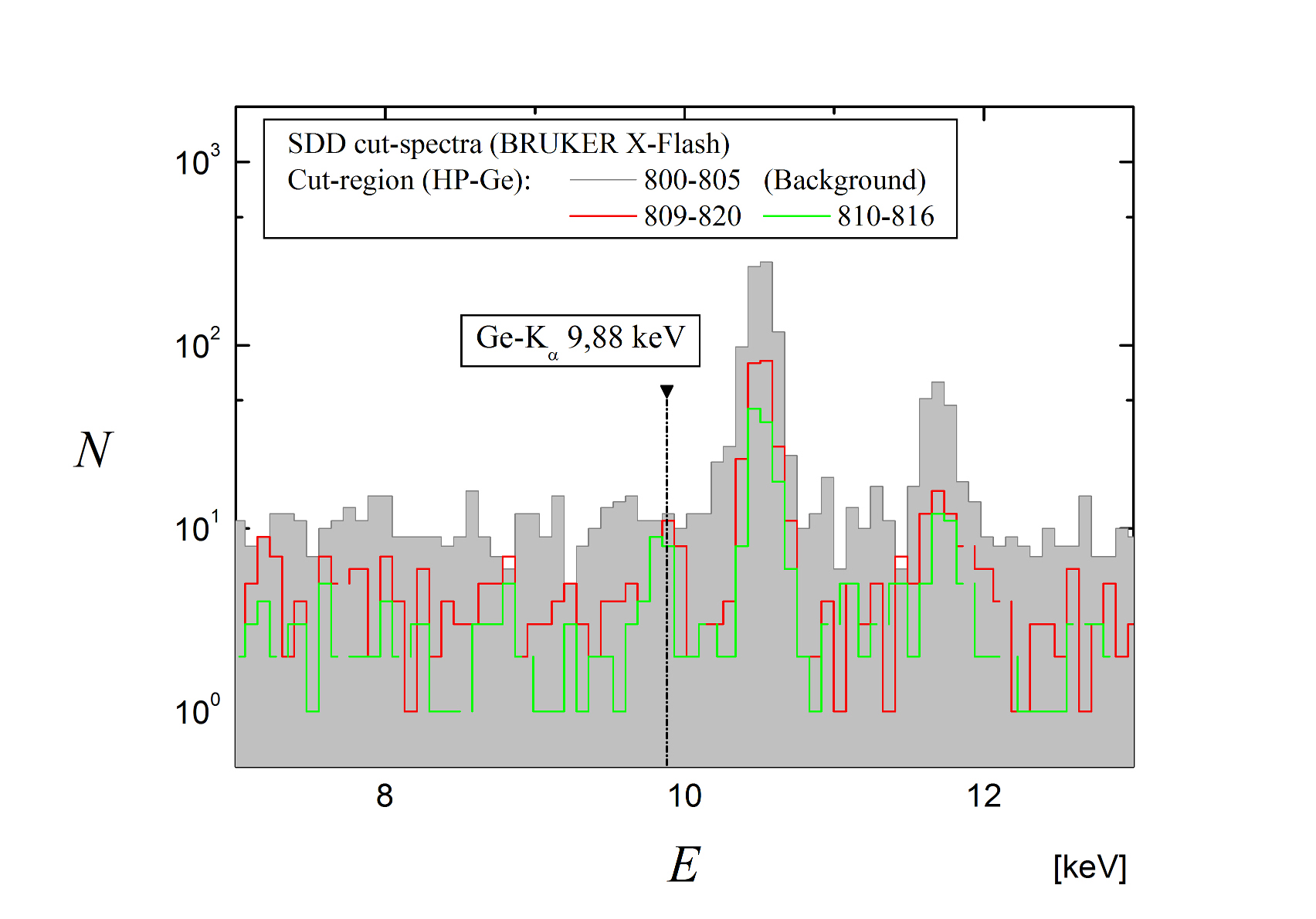}
\caption{\textit{Left:} The $\gamma$-spectrum shows the weak 592.93\,keV line (Figure \ref{pic:decscheme} $\gamma_1$) from the de-excitation of the \gess-nucleus in the cut-region which is overlapped in a peak-triplet with the strong lines at 559.101\,keV and 563.178\,keV from the $\beta^-$-channel (Figure \ref{pic:decscheme} $\gamma_{2/3}$). The cut-efficiency is illustrated by the area of the gaussian (green) which depends on the boundaries of the cut (blue). \textit{Right:} The resulting X-ray cut-spectra is dominated by the Se-K$_{\alpha1/2}$-lines as spectral background as shown with the cut on the region of $\gamma_{2}$ (grey). The expected Ge-K$_{\alpha1/2}$-lines show clearly up in the cutregions of the 592.93\,keV-line (red, green) in the cut-spectra.}
\label{pic:BP-CUT}
\end{figure*}

The potential weak FEP of the 592.93\,keV $\gamma$-line (Figure \ref{pic:decscheme}, $\gamma_1$) is not directly visible due to an overlap with the two strong FEPs of the lines at 559.101\,keV and 563.178\,keV from the $\beta^-$-channel (Figure \ref{pic:decscheme} $\gamma_{2/3}$). Nevertheless, a cut-region in the HP-Ge-spectra was defined around the $\gamma_{1}$-peak region at 592.93\,keV, whereas all coincident events were reconstructed in the corresponding X-ray cut-spectra (Figure \ref{pic:BP-CUT}). The upper and lower limit of the cut was varied to get an optimal signal-background-ratio for the Ge-K$_{\alpha}$-line in the corresponding cut-spectra. The broader the cut-region in the $\gamma$-spectra, the higher the number of events of the neighboring peaks from the $\beta^-$-channel and the higher the background in the X-ray-cut-spectra which is dominated by particles, that are coming from the  \asss-$\beta^-$-channel (Figure \ref{pic:BP-CUT} \textit{right}). With respect to this background the cut-region was chosen as big as necessary but not bigger than reasonable. Accordingly the cut does not cover all events which were counted in the overlaped 592.93\,keV FEP in the $\gamma$-spectra. The cut-efficiency

\begin{equation}
\varepsilon_{\mathrm{cut}} =  \left.\frac{A_{\mathrm{cut}}}{A_{\mathrm{gauss}}} \right|_{FWHM(E_{\gamma})} 
\label{eq31}
\end{equation}

at a given FWHM is calculated from the peak-fraction which is covered by the cut using the areas of the gaussian peaks of the cut $A_\mathrm{cut}$ and the full peak area $A_\mathrm{gauss}$ (Figure \ref{pic:BP-CUT}). Based on the FEP-area in the X-ray-cut-spectra $A_\mathrm{coinc}$ (Table \ref{Table5}) the coincident counted events 

\begin{eqnarray}
N_\mathrm{coinc} & = & \frac{A_\mathrm{coinc}}{\varepsilon_\mathrm{cut} }
\label{eq32}
\end{eqnarray}

are directly accessible from the experiment with respect to the cut-efficiency (Eq. \ref{eq31}). This number of coincident counted events
\begin{eqnarray}
\label{eq33}
N_\mathrm{coinc} & = & Z_\mathrm{total} \cdot p_\mathrm{EC} \cdot p_\mathrm{1^{st}} \cdot \nu_\mathrm{rel} \left( \gamma_i \right) \nonumber \\
 & &  \cdot  \omega_\mathrm{K} \left( \asss \right) \cdot \left( \nu_{\alpha1}+\nu_{\alpha2} \right)  \nonumber \\
 & &  \cdot  \varepsilon_\mathrm{SDD} \left( E_{Ge-K_{\alpha}} \right) \cdot \varepsilon_\mathrm{HP-Ge} \left( E_\gamma \right)
\end{eqnarray}

in the HP-Ge-detector and SDD with the corresponding FEP-efficiencies $\varepsilon_{i}$ is given by the number of total decays $Z_\mathrm{total}$, the K-capture-fraction $\omega_\mathrm{K}\left(\asss\right)$, the emission-probabilities of the Ge-K$_{\alpha}$-lines $\left(\nu_{\alpha1}+\nu_{\alpha2}\right)$ the $\beta^{+}$/$\beta^{-}$ branching-ratio $p_\mathrm{EC}$, the branching-ratio of the EC-channel itself $p_\mathrm{1^{st}}$ and the emission-probability $\nu_\mathrm{rel}\left(\gamma_i\right)$ of the coincident line $\gamma_i$ with the energy $E_{\gamma_i}$. With the use of the total counted Ge-K$_{\alpha}$ peak events in the SDD (Table \ref{Table2})

\begin{eqnarray}
N_{Ge-K_{\alpha}} & = & Z_{total}\cdot p_{EC} \cdot \omega_K\left(\asss\right) \nonumber \\
                  &   & \cdot \left(\nu_{\alpha1}+\nu_{\alpha2}\right) \cdot \varepsilon_{SDD}\left(E_{Ge-K_{\alpha}}\right)
\label{eq34}
\end{eqnarray}

which is also directly accessable from the X-ray-spectra, the branching-ratio of the EC-channel for the first excited state of \gess

\begin{eqnarray}
p_{1^{st}} & = & \frac{N_{coinc}}{N_{Ge-K_{\alpha}}} \cdot \frac{1}{\nu_{rel}\left(\gamma_i\right) \cdot \varepsilon_{HP-Ge}\left(E_\gamma i\right)}
\label{eq35}
\end{eqnarray}

results from Equation \ref{eq33} and \ref{eq34}. To evaluate a systematic uncertainty $ \Delta N_\mathrm{coinc,sys}^\mathrm{Ge-K_{\alpha}} $ for the cut the analysis was repeated with coincidence-spectra from different cut-regions with variation of the boundaries of the gaussian-fit (Figure \ref{pic:BP-CUT} \textit{left}). 
The higher the coverage of the lower boundary, the higher is the increase of the background in the X-ray cut-spectra. To prevent this huge background for the small number of coincident peak-events, the lower bound was constrained at channel 809, which is low enough for a good cut-coverage at a reasonable background-level. For a cut-coverage variation the right bound was varied.


\begin{table*}[!htb]
\caption{The coincident Ge-K$_{\alpha}$-$\gamma_1$ events $N_\mathrm{coinc}^{Ge-K_{\alpha}}$ were derived from Ge-K$_{\alpha}$ peak-fits in the X-ray-cut-spectra with the area $A_\mathrm{coinc}$ for different boundaries of the cut (Figure \ref{pic:BP-CUT})}
\label{Table5} 
\begin{tabularx}{1.0\linewidth}{p{0.2\linewidth}  p{0.1\linewidth}  p{0.1\linewidth}  p{0.1\linewidth} p{0.1\linewidth} p{0.1\linewidth}  p{0.1\linewidth}  p{0.1\linewidth}} \hline \hline
 &   &  &  &  &  &  &  \\
cut [channel] & $A_\mathrm{coinc}$  & $\Delta A_\mathrm{coinc,stat}$ & $\Delta A_\mathrm{coinc,sys}$ &  & $N_\mathrm{coinc}^{Ge-K_{\alpha}}$  & $\Delta N_\mathrm{coinc,stat}^{Ge-K_{\alpha}}$ & $\Delta N_\mathrm{coinc,sys}^{Ge-K_{\alpha}}$ \\ 
 &   &  &  &  &  &  &  \\ \hline
 &   &  &  &  &  &  &  \\
809-820 & 19.96 & 6.52 & 0.93 &  & 3257 & 1063 & 238  \\
810-816 & 17.64 & 5.22 & 0.59 &  & 3310 & 980 & 204  \\ \hline
 &   &  &  &  &  &  &  \\
weighted mean &  &  &  &  & 3286 & 720 & 221 \\
 &   &  &  &  &  &  &  \\ \hline
\end{tabularx}
\end{table*}

Taking the weighted mean of the two cuts in Table \ref{Table5} the branching of the EC-channel itself to the 1$^\mathrm{st}$ excited state of \gess was calculated to

\begin{eqnarray}
p_{1^{st}}  =  132\%\pm48(stat.)\%\pm9(sys.)\%.
\label{eq36}
\end{eqnarray}

As a physical reasonable result an upper and lower limits 

\begin{eqnarray}
p_{1^{st}} \,& \geq &\, 75\% (68\% C.L.) \nonumber \\
p_\mathrm{ground}\,& < &\, 25\% (68\% C.L.). 
\label{eq37}
\end{eqnarray}

for the EC into the first excited and the ground state of \gess were calculated.

\subsection{Uncertainties}

Statistical and systematic errors were calculated based on the uncertainties of the measured quantities $\Delta x_i$ according to the gaussian-error-propagation

\begin{eqnarray}
\Delta F_\mathrm{sys./stat.}=\sqrt{\sum_{i=1}^N \left|\frac{\partial}{\partial x_i}F\left(x_1 ,x_2,..,x_N \right) \cdot \Delta x_i \right|^2}
\label{eq38}
\end{eqnarray}

independently from each other and summed up to the total error

\begin{eqnarray}
\Delta F=\Delta F_\mathrm{sys}+\Delta F_\mathrm{stat}.
\label{eq39}
\end{eqnarray}

A systematic uncertainty for any peak fit  was accounted with a set of $M$ approximations with different boundaries. The results were merged in a weighted mean

\begin{eqnarray}
\overline{x} = \frac{\sum_{i=1}^M x_i \cdot \sigma_i^{-2}}{\sum_{i=1}^M \sigma_i^{-2}}
\label{eq40}
\end{eqnarray}

with the corresponding uncertainty

\begin{eqnarray}
\overline{\sigma} = \sqrt{ \frac{\sum_{i=1}^M 1}{\sum_{i=1}^M \sigma_i^{-2}}}.
\label{eq41}
\end{eqnarray}

Moreover, the influence of systematic errors, which are caused by various sources like the powder density and positioning error, were quantified with the use of a Monte-Carlo-simulation. The conversion-factors account for this influences with the uncertainty for the respective values (Equation \ref{eq1} \& \ref{eq2}). As the effect of the K$_{\alpha}$-lines on the above discussed systematic influences is correlated, the error of the total branching-ratio is rather small using the analysis based on the Se-K$_{\alpha}$-line.

\section{Results}

Besides the dependance on the $Q$-value for the probability of a $\beta$-decay and the half-life $t_{1/2}$ of the respective decay-channel, there is another strong dependance on the change of angular momentum and parity $\Delta J^\pi$ for the transition \cite{suhonen07}. The log$ft$-value

\begin{eqnarray}
\mathrm{log}ft=\mathrm{log}\left(f_x \cdot t_{1/2} \right).
\label{eq50}
\end{eqnarray}

accounts for this dependency with the Fermi-function $f_x$. It describes the strength of the transition and gives access to a classification of the respective $\beta$-decay-channel. For a radionuclide with the half-life $T_{1/2}$ the partial half-life

\begin{eqnarray}
t_{1/2}^i = \frac{T_{1/2}}{p_i}
\label{eq51}
\end{eqnarray}

of the $i$-th decay-channel is derived from its branching-ratio $p_i$. Although for the present experiment a ground state-EC was not detected the limits are derived for the log$ft$-values from the branching-ratio-limits (Equation \ref{eq37}).

The Fermi-phase-space-integral values for the unique- and non-unique first-forbidden EC are derived from \cite{govemartin71}. Whereas in the first case
$f_{1u}$ is directly derived, \cite{suhonen07} suggests to use the value of the allowed EC/$\beta^{+}$-decay $f_{0}$ as Fermi-phase-space-integral for the non-unique first-forbidden EC $f_{1}$.

\begin{figure*}[!htb]
\includegraphics[width=0.95\columnwidth]{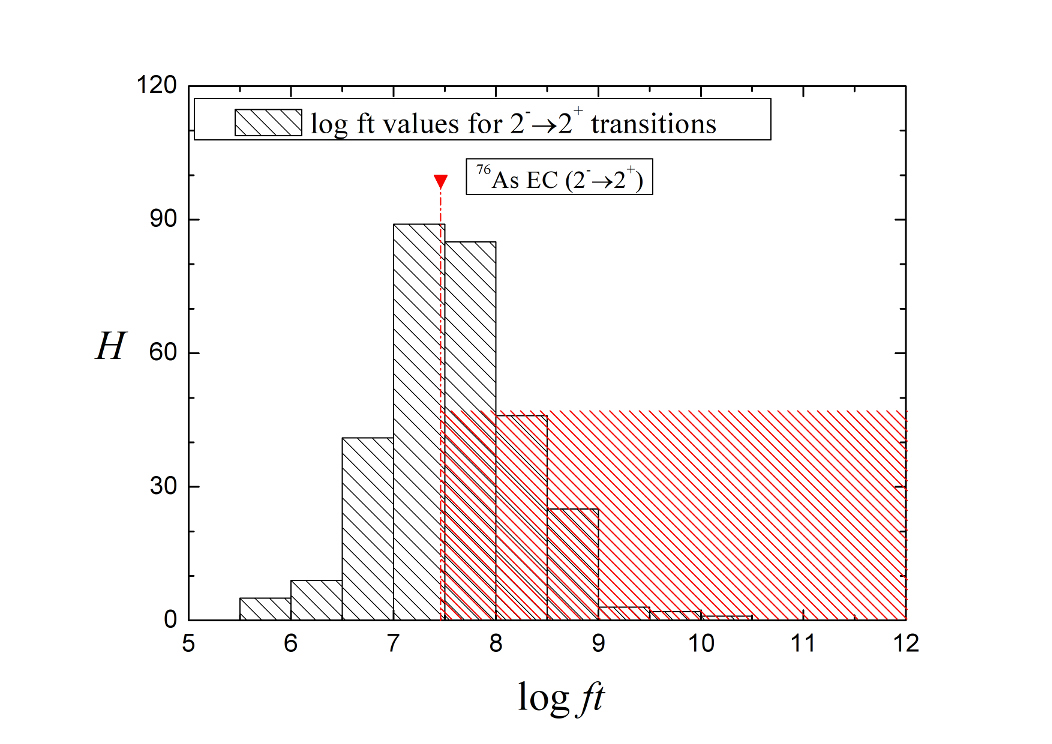}
\includegraphics[width=0.95\columnwidth]{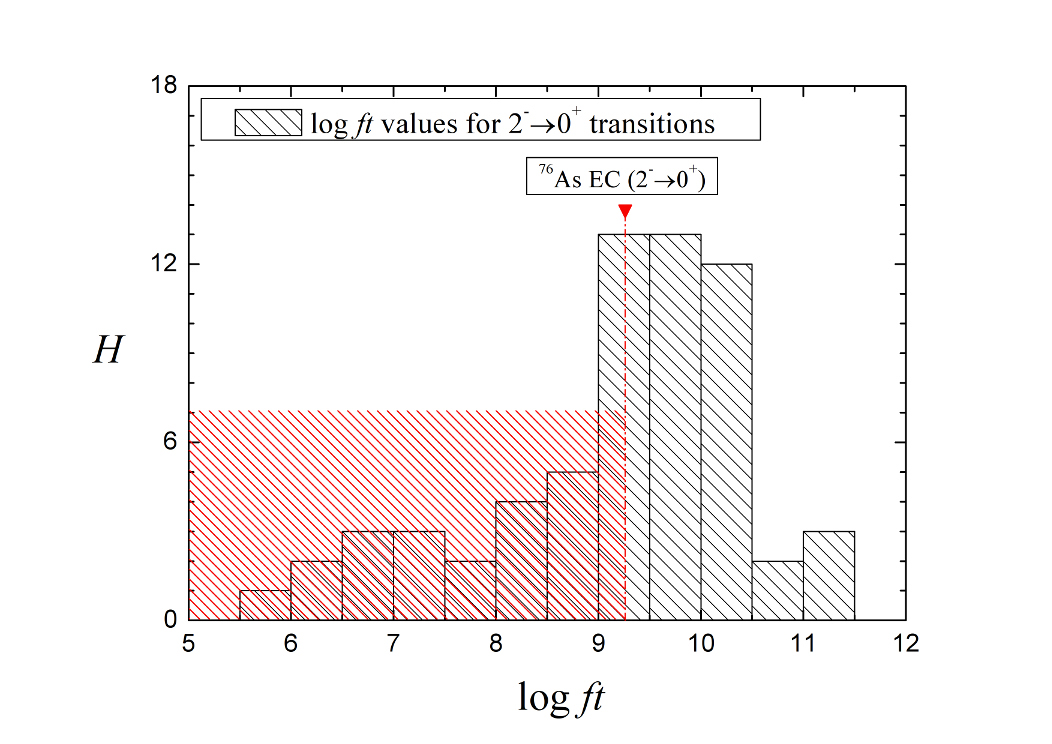}
\caption{For a comparison with evaluated data, log$ft$ values from ENSDF \cite{ENSDF} for similar transitions were analysed. The upper and lower limits for log$ft$-values of the \asss-EC-channels in the ground-state (\textit{left}) and the first excited-state (\textit{right}) define exclusion regions (red).}
\label{pic:logft}
\end{figure*}

Assuming the partial half-lives for the transitions the log$ft$-limits result according to Table \ref{Table6}. The values derived from the present experiment are opposed to ENSDF-data \cite{ENSDF} for comparable $2^{-}\longrightarrow 2^{+}$ first forbidden non-unique  (Figure \ref{pic:logft} \textit{left}) and $2^{-}\longrightarrow 0^{+}$ first forbidden unique (Figure \ref{pic:logft} \textit{right}) transitions. Both values and exclusion regions show a good agreement with the present data.

\begin{table*}[!htb]
\caption{Using the Fermi phase-space-integrals from \cite{govemartin71} the log$ft$-limits were derived from the branching-ratio-limits of this measurement for the two EC-transitions of \asss. In comparison with the calculated values for the branching-ratio with the use of log$ft$-values from the classification of the respective transitions, the measured limits show up reasonable.}
\label{Table6} 
\begin{tabular*}{1.0\linewidth}{p{0.2\linewidth}  p{0.1\linewidth}  p{0.05\linewidth} p{0.1\linewidth}  p{0.1\linewidth} p{0.1\linewidth} p{0.05\linewidth}  p{0.1\linewidth}  p{0.1\linewidth}} \hline \hline
 &  &   &  &  &  &  &  &  \\
 &  &    \multicolumn{4}{c}{measured}  & \multicolumn{3}{c}{calculated}  \\

 Transition & $f_x$ & & $t_\mathrm{1/2}$ [$10^{9}$ s] & log$ft$ & $p_\mathrm{i}$ & & log$ft$ & $p_\mathrm{i}$ \\ 
 &  &   &  &  &  &  &  &  \\
 & \cite{govemartin71}  & &  &  &  &  & \cite{TMK92} &  \\
 &   &  &  &  &  &  &  &  \\ \hline
 &   &  &  &  &  &  &  &  \\

$2^{-}\longrightarrow 2^{+}$  & 0.062  & & $<0.47$ & $<7.46$ & $\geq75\,\%$ &  & $7.5 \, \pm \, 1.5$ & $68\% \pm 83\%$ \\
$2^{-}\longrightarrow 0^{+}$  & 1.32   & & $>1.39$ & $>9.26$ & $<25\,\%$ &  & $8.5 \, \pm \, 0.7$ &  $32\% \pm 83\%$ \\

 &   &  &  &  &  &  &  \\ \hline
\end{tabular*}
\end{table*}

Based on the log$ft$-values as given by \cite{TMK92} the branching-ratio of the EC-channel was estimated. 
Also if the given uncertainties result in a huge effect for the branching-ratio the limits from the present experiment show up reasonable in comparison with the estimation.

\section{Summary and conclusions}

In the present experiment the electron-capture of \asss has been observed for the first time. The total $\beta^{+}$/$\beta^{-}$ branching-ratio was determined by the use of two different methods. Whereas both methods result in very similar values, the first method of analysis comes up with smaller systematic uncertainties. Therefore, the main reason is that the analysis is based on a dataset of one detector only and one dominating uncertainty for the FEP-efficiency is eliminated. However the matching results validate the values which are derived from both analysis-methods. Within their uncertainties both values agree with the evaluated limit of J. Scobie \cite{sco57}. Moreover, a confirmation is given for relative X-ray-intensities and K-capture fractions, which are mostly based on theoretical origin.\\
The EC branches into the first forbidden non-unique and unique transition, which branching ratio could not been quantified in this experiment. Nevertheless, lower and upper limits were derived for the branching into the first excited- and into the ground state of \gess. It is shown that the method works nicely also if the main-limitation in the present experiment is given by poor statistics in the cut-spectra.\\
As main-result of this paper the log$ft$-limits were derived from the branching ratios for both EC-channels. In comparison with evaluated data both limits are reasonable.

\section{Acknowledgement}
We thank the team of the reactor AKR-II from Prof. W. Hansen at TU Dresden for the activation of the sample and the radiation-physics-group of PD J. Henniger (ASP/IKTP) for the supply of the MC-code AMOS and hardware support. Many thanks to Prof. J. Suhonen for helpfull correspondences. This work is supported by the BMBF grant 05A08OD1.

\bibliography{\jobname}

\begin{thebibliography}{10}

\bibitem{singh95}
B.Singh.
\newblock {\em \NDS}, 74:63, 1995.

\bibitem{ame12}
M.Wang et~al.
\newblock {\em \CPC}, 36:1603, 2012.

\bibitem{mor71}
P.~K.Kuroda N.E.~Morcos, T.E.~Ward.
\newblock {\em \NPA}, 171:647, 1971.

\bibitem{mcm71}
D.K McMillan and B.D.Pate.
\newblock {\em \NPA}, 174:604, 1971.

\bibitem{cam98}
S.P. de~Camargo~et al.
\newblock {\em \ARI}, 49:997, 1998.

\bibitem{mims51}
H.Halban W.Mims.
\newblock {\em Proc.Phys.Soc. A}, 64:311–312, 1951.

\bibitem{sco57}
J.~Scobie.
\newblock {\em \NP}, 3:465, 1957.

\bibitem{xcom12}
S.M.~Seltzer et~al.
\newblock Xcom: Photon cross section database (version 1.5) [online].
  gaithersburg: National institute of standards and technology.
\newblock 2012.

\bibitem{akr}
{\em "Technical Description and Procedure of Operation for the Reactor Facility
  AKR-2"}, 2015.

\bibitem{gun77}
R.~Gunnink.
\newblock {\em Nuclear Instruments and Methods}, 143:145, 1977.

\bibitem{toi98}
R.~Firestone.
\newblock Table of isotopes, 8th edition.
\newblock 1998.

\bibitem{henn05}
J.~Henniger.
\newblock {\em Anleitung f\"ur die Benutzung des Strahlungstransportcodes AMOS
  1.0}, 2005.

\bibitem{hanson68}
R.J.~Hanson et~al.
\newblock {\em Nuclear Physics A}, 115:641, 1968.

\bibitem{logft01}
{\em LogFT}, 2001.

\bibitem{suhonen07}
J.~Suhonen.
\newblock {\em "From Nucleons to Nucleus"}.
\newblock 2007.

\bibitem{govemartin71}
M.J.~Martin N.B.~Gove.
\newblock {\em Nuclear Data Tables}, 10:205--317, 1971.

\bibitem{ENSDF}
M.R. Bhat.
\newblock {\em "Evaluated Nuclear Structure Data File (ENSDF)", Nuclear Data
  for Science and Technology}.
\newblock Springer-Verlag, Berlin, Germany, 1992.

\bibitem{TMK92}
T.~Mayer-Kuckuk.
\newblock {\em "Kernphysik" 5. \"uberarbeitete und erg\"anzte Auflage}.
\newblock Verlag B.G. Teubner, Stuttgart, 1992.

\end{thebibliography}

\end{document}